\documentclass[prb,twocolumn,amsmath,amssymb,amsfonts,superscriptaddress,showpacs]{revtex4-1}
\usepackage{graphicx}
\usepackage{subfigure}
\usepackage{dcolumn}
\usepackage{bm}
\usepackage{color}
\usepackage{amsmath}
\usepackage{float}
\usepackage{subfig}
\usepackage{xcolor}
\usepackage[export]{adjustbox}
\usepackage{xspace}
\usepackage{maybemath}

\def\bra#1{\mathinner{\langle{#1}|}}
\def\ket#1{\mathinner{|{#1}\rangle}}
\def\braket#1{\mathinner{\langle{#1}\rangle}}

\newcommand{\etal}{\textit{et al.\/}}

\newcommand{\eg}{e.\,g.}

\newcommand* \tk{{\bf k}}

\begin{document}

\title{Charge self-consistency in density functional theory + dynamical mean field theory: \tk-space reoccupation and orbital order}

\author{Sumanta Bhandary}
\email{bhandary@ifp.tuwien.ac.at}
        \affiliation{Institute of Solid State Physics, TU Wien, 1040 Wien, Austria}
\author{Elias Assmann}
        \affiliation{Institute of Solid State Physics, TU Wien,  1040 Wien, Austria}
        \affiliation{Institute of Theoretical and Computational Physics, Graz University of Technology, Petersgasse 16, 8010 Graz, Austria}
\author{Markus Aichhorn}
        \affiliation{Institute of Theoretical and Computational Physics, Graz University of Technology, Petersgasse 16, 8010 Graz, Austria}
\author{Karsten Held}
        \affiliation{Institute of Solid State Physics, TU Wien,  1040 Wien, Austria}
        
\begin{abstract}
We study effects of charge self-consistency within the combination of density functional theory (DFT; Wien2k)
with  dynamical mean field theory (DMFT; w2dynamics) in a basis of maximally localized Wannier orbitals.  Using the example of two  cuprates, we demonstrate that even if there is only a single Wannier orbital with fixed filling, a noteworthy charge redistribution can occur. This effect stems from a reoccupation of the Wannier orbital in \tk-space when going from the single, metallic DFT band to the split, insulating Hubbard bands of DMFT.  We analyze another charge self-consistency effect beyond moving charge from one site to  another: the correlation-enhanced orbital polarization in a freestanding layer of SrVO$_3$.

\end{abstract}

\maketitle

\section{Introduction}

Density functional theory (DFT)\cite{dft1,dft2} is highly successful in predicting various material properties such as  crystal structures, ionization energies, electrical, magnetic and vibrational properties. Indeed DFT is the {\em de facto} standard for calculating materials'  physical properties. But 
even the best approximations for the DFT exchange-correlation functional fail to describe one class of materials, known as strongly correlated systems.  In these materials, the interaction between electrons is insufficiently screened to be amenable to the available functionals.  One might add a static Coulomb correction within the so-called DFT+U formalism.\cite{ldau} This often yields an improved description, in particular  of strongly correlated insulators,  but it has its own limitations: DFT+U is essentially a Hartree-Fock-like treatment with a single-Slater-determinant  ground state. In this situation, the  energy cost of the Coulomb interaction can only be avoided by symmetry breaking, which is hence largely overestimated.

Dynamic, albeit local, correlations can be taken into account by dynamical mean field theory (DMFT) \cite{Metzner89a,Georges92a,Georges96a} which has been merged with DFT for realistic calculations of correlated materials.\cite{Anisimov97a,Lichtenstein98a,dmft1, dmft2} Here electrons can stay on or leave lattice sites dynamically so as to greatly suppress  double occupation and the cost of the Coulomb interaction, even in a paramagnetic phase without any symmetry breaking. If one has a three dimensional material at elevated temperatures, say room temperature, and if there is no magnetic or other phase transition close-by,\cite{Rohringer2011}  these local DMFT correlations prevail. Already the first applications  showed that DFT+DMFT  well describes transition metals \cite{Lichtenstein98a},
their oxides \cite{Held01a}, and f-electron systems \cite{SAVRASOV,Held01b}.

These early calculations were so-called ``one-shot''. That is following a DFT calculation, the
relevant correlated orbitals and the corresponding single-particle Hamiltonian were identified. This DFT Hamiltonian was supplemented by  local Coulomb interactions for the $d$- or $f$-orbitals and solved with DMFT. Physical properties such as the spectral function, susceptibility or magnetization were calculated from this ``one-shot'' DMFT solution.

Since the DMFT correlations change the site and orbital occupation and consequently  the charge density, a natural next step is to do a ``charge self-consistent'' (CSC) DFT+DMFT  calculation.\cite{Savrasov04,Minar05,millis_csc,Frank,Pouroskii07,Aichhorn2011,haule2010} That is, from the DMFT Green's function, a new charge distribution is calculated which in turn serves as input for the  DFT potential. This leads to a new DFT Kohn-Sham Hamiltonian, subsequently a new DMFT Green's function etc.\ This cycle is repeated until convergence. While it has been pointed out in the literature \cite{Savrasov04,Minar05,millis_csc,Frank,Pouroskii07,Aichhorn2011,haule2010} how the DMFT spectral function, the  double counting and the $d$ ($f$) energy level changes due to CSC for specific materials where charge is moved from one site to another, little attention has been paid to the redistributed charge itself, its spatial arrangement, and more.

The aforementioned change of double counting and $d$ level shift can be understood as follows: In a typical situation, say for a transition metal oxide, the dominant $d$ states crossing the Fermi energy  have some oxygen $p$ admixture; conversely, the oxygen states below the Fermi level have some $d$ contribution. Including electronic correlations in a so called $d$+$p$  DMFT calculation will reduce the $d$ occupation somewhat, and increase the $p$ occupation on the oxygen sites.
In the next DFT step, the larger $p$ occupation will increase the $p$ (Hartree) energy and decrease the $d$ (Hartree) energy. This counteracts the first shot DMFT to have less $d$ and more $p$ electrons, dampening the charge redistribution of the ``first-shot'' DFT+DMFT.

In this paper we study effects of CSC beyond this gross effect of a $p$-$d$
orbital and site reoccupation. In Section \ref{sec:method}, we recapitulate the CSC DFT+DMFT approach and outline our implementation thereof. 
In Sections  \ref{sec:SCTO} and  \ref{sec:HBCO}, we show that even in a single-orbital, $d$-only DMFT calculation,  there is a charge redistribution akin the $d$-$p$ reoccupation effect mentioned above. This runs counter to the naive expectation that there can be no charge redistribution in this situation since the number of electrons in the single, predominately $d$-like orbital centered around the 
transition metal site is fixed. 
The two materials studied, where a restriction to a single $d$ band is justified, are
 Sr$_2$CuTeO$_6$ and HgBa$_2$CuO$_4$ 
in Section  \ref{sec:SCTO} and  \ref{sec:HBCO}, respectively.
In Section 
 \ref{sec:SVO},  we study the effect of correlation-induced orbital order on the charge
redistribution and self-consistent DFT+DMFT results. Specifically, we consider an ultra-thin layer of the cubic perovskite material SrVO$_3$, where breaking of the cubic symmetry 
stabilizes the in-plane $xy$ orbital  against the $xz$ and $yz$ orbitals. This orbital ordering is strongly enhanced in DMFT because of electronic correlations. Finally, 
 Section 
 \ref{sec:summary} summarizes our main findings.


\section{Methodology}
\label{sec:method}

We now present the formalism and our implementation of CSC DFT+DMFT which is in a basis of maximally localized Wannier functions (MLWF).  For these, the measure of localization introduced by Marzari and Vanderbilt\cite{wan1} is the spread in real space. This provides for a very flexible approach portable to any bandstructure method. Moreover, the methodology allows bond-centered or molecular Wannier functions.  Our starting point is the wien2wannier\cite{wien2wannier} interface between Wien2k\cite{w2k} and Wannier90 \cite{wanrev}, and the w2dynamics\cite{w2d} continuous-time quantum Monte Carlo \cite{CTQMC} DMFT implementation.

We combine and extend these methods to include CSC. Let us, for the sake of completeness and given the sparse presentation in the literature,  recapitulate here the CSC DFT+DMFT approach, and discuss the peculiarities of our implementation. Readers only interested in the physical applications and effects of CSC can safely skip the rest of this Section.


The CSC DFT+DMFT method relies on the simultaneous convergence of two local observables: the electronic density as the central quantity of DFT and the local Green's function as the central quantity of DMFT. Both mutually affect each other in the CSC cycle.
The charge density at position $\mathbf r$ is given by
\begin{equation}
\rho({\bf r}) = \frac{1}{\beta}\sum_n G({\bf r},{\bf r};i\omega_n)e^{i\omega_n0^+},
\end{equation}
while the local DMFT Green's function is
\begin{equation}
  G_{mm'}(i\omega_n)=\! \int\! d{\bf r}d{\bf r'} \chi^*_m({\bf r})
  \chi_{m'}({\bf r'})G({\bf r},\!{\bf r'}\!;\!i\omega_n)
\end{equation}
in the basis of localized Wannier orbitals $\chi_m$.  Here, $m, m'$ enumerate orbitals on a site, $\beta$ is the inverse temperature, and the factor $e^{i\omega_n0^+}$ ensures the convergence of the sum over Matsubara frequencies $\omega_n=(2n+1)\pi /\beta$.

In both expressions there appears the full Green's function of the solid, which can be written as
\begin{equation}
\begin{aligned}
G({\bf r},{\bf r'};i\omega)=\bra{{\bf r}}[i\omega_n+\mu+\frac{\nabla^2}{2}-\hat{V}_{KS}-\Delta\hat{\Sigma}]^{-1}\ket{{\bf r'}},
\end{aligned}
 \end{equation}
with $-\frac{\nabla^2}{2}$, $V_{KS}$, and $\mu$ being the kinetic energy operator, Kohn-Sham (KS) effective potential, and chemical potential, respectively.  The effective local self-energy $\Delta\hat{\Sigma} = \hat{\Sigma} - \hat{\Sigma}_{dc}$ is determined from the DMFT self-energy $\hat{\Sigma}$ by subtracting a double counting correction term $\hat{\Sigma}_{dc}$, which, as far as possible, accounts for electronic correlations already included in DFT.  The KS potential depends on ${\bf r}$ and consists of the external potential $V_{ext}$ due to the nuclei, a Hartree potential $V_{H}$ describing part of the  electron-electron Coulomb repulsion and an exchange-correlation potential $V_{xc}$. The latter is obtained here within the generalized gradient approximation (GGA)\cite{gga}, but other functionals are possible as well, e.g. hybrid functionals to improve on the exchange contribution.

\begin{figure*}[t]
 \centering
 \includegraphics[width=\textwidth]{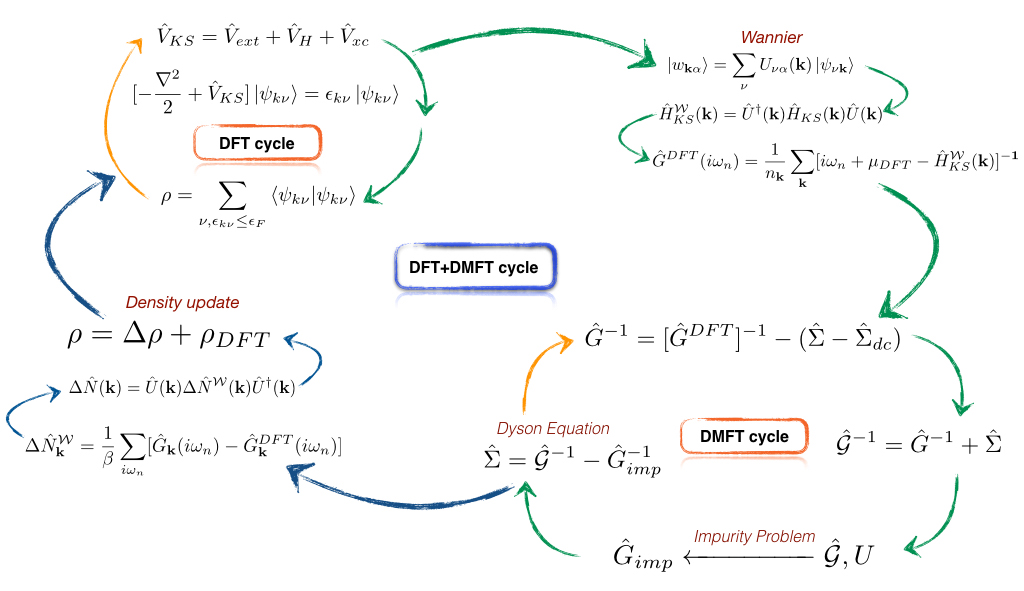}
\caption{\label{cycle} Schematic representation of the DFT+DMFT approach. In a non-CSC or ``one shot'' DFT+DMFT calculation, the DFT Hamiltonian is not updated and both the DFT and DMFT cycle close individually, i.e., we have the orange and green arrows in the schematic, but not the blue ones. In a CSC DFT+DMFT, neither DFT nor  DMFT is iterated  individually.  Instead, both of them are closed together,  i.e., we have the green and blue arrows in the schematic, but not the orange ones.}
\end{figure*}

 In DFT, the effective potential is obtained from a charge self-consistent procedure, shown in the upper left part of Fig \ref{cycle}. This DFT cycle starts with an initial choice for the electron density, from which the effective potential $V_{KS}$  is constructed. Incorporating $V_{KS}$, the Kohn Sham equation is solved to obtain a new density and so forth until convergence. The DFT cycle closes with a converged charge density and provides a reasonable electronic structure as a starting point for  DMFT calculations.

There is however an important step between DFT and DMFT, identifying 
a localized basis (upper right part in  Fig \ref{cycle})  since DMFT treats only local correlations. To this end, 
we employ  Wannier functions that are constructed by Fourier transform of the DFT Bloch waves $\ket{\psi_{\nu {\bf k}}}$:
\begin{equation}\label{wan0}
\begin{aligned}
\ket{w_{\alpha\bf R}}=\frac{\Omega}{(2\pi)^3}\int_{BZ} d{\bf k}~e^{-i{\bf kR}}\sum_{\nu=1}^{ {\mathcal C}}U_{\nu\alpha}({\bf k})\ket{\psi_{\nu {\bf k}}}.
\end{aligned} 
\end{equation} 

Here, $\hat{U}({\bf k})$ is a unitary matrix, $\Omega$  denotes the volume of the unit cell, $\nu$ and $\alpha$ are the band indices of the Bloch waves and Wannier functions, respectively. We assume here that we can restrict ourselves to a band window with only   ${\mathcal C}$ Bloch waves.
 In the scheme of maximally localized Wannier functions\cite{wan1},   $\hat{U}({\bf k})$ is obtained by minimizing the spread of the Wannier functions.

 Eq.\ (\ref{wan0}) works for isolated bands. However, in most cases, the target bands are $``$entangled$"$ with further bands at least at some \tk-points.  These additional bands might be less important for the physics but need to be projected out by a so-called $``$disentanglement" procedure. At each ${\bf k}$-point, there is  a  set of  ${\mathcal C}^o({\bf k})$ Bloch functions  which is larger than or equal to the number of target bands, i.e.,  ${\mathcal C}^o({\bf k)} \ge {\mathcal C}$.  The disentanglement transformation takes the form
\begin{equation}\label{wan1}
\begin{aligned}
\ket{w_{\alpha\bf R}}=\frac{\Omega}{(2\pi)^3}\int\limits_{BZ} d{\bf k}~e^{-ik{\bf R}}\sum_{\nu'=1}^{ {\mathcal C}} \sum_{\nu=1 }^{ {\mathcal C^o({\mathbf k})}}V_{\nu\nu'}({\bf k}) U_{\nu'\alpha}({\bf k})\ket{\psi_{\nu {\bf k}}}.
\end{aligned}
\end{equation}
Here, the band index $\nu$  belongs to the $``$outer window" with ${\mathcal C}^{o}({\bf k})$ Bloch wave functions, while $\nu',\alpha$ label the $\mathcal C$ target bands. Hence, $\hat{V}({\bf k})$ is a rectangular   ${\mathcal C}^{o}({\bf k}) \times {\mathcal C}$  matrix. A Fourier transformation of $\ket{w_{\alpha\bf R}}$ leads to the Wannier orbitals in \tk-space whose occupation will be at the focus of the physics discussed below:
\begin{equation}\label{wan2}
\begin{aligned}
\ket{w_{\alpha{\bf k}}}=\sum_{\bf R}e^{ik{\bf R}}\ket{w_{\alpha\bf R}}=\sum_{\nu'\nu}V_{\nu\nu'}({\bf k})U_{\nu'\alpha}({\bf k})\ket{\psi_{\nu {\bf k}}}
\end{aligned}
\end{equation}

The Hamiltonian in Wannier space ${\mathcal W}$ is defined in terms of the 
$\ket{w_{{\alpha}{\bf k}}}$ and obtained by an unitary transformation for isolated bands and with an additional projection (``downfold'') in case of entangled bands, i.e.,
\begin{eqnarray}\label{ham}
\hat{H}^{\mathcal{W}}_{KS}({\bf k})&=&\hat{U}^{\dagger}({\bf k})\hat{H}_{KS}({\bf k})\hat{U}({\bf k}),\\
\hat{H}^{\mathcal{W}}_{KS}({\bf k})&=&\hat{U}^{\dagger}({\bf k})\hat{V}^{\dagger}({\bf k})\hat{H}_{KS}({\bf k})\hat{V}({\bf k})\hat{U}({\bf k})
\end{eqnarray}
respectively.

In DMFT, this Hamiltonian is now supplemented with the local Coulomb interactions,
and the lattice problem defined this way is mapped onto an 
auxiliary impurity problem which is solved self-consistently.\cite{Georges92a,Georges96a}
The non-interacting Green's function  $\hat{\mathcal{G}}(i\omega_n)$
of the impurity problem can be considered as a  dynamical mean field.
The DMFT algorithm (see the lower right part of Fig.~\ref{cycle}) consists of:

(i) Applying the lattice Dyson equation for the local interacting Green's function $\hat{G}(i\omega_n)$
\begin{eqnarray}\label{locg}
\begin{aligned}
\hat{G}(i\omega_n)=\frac{1}{n_{\bf k}}\sum_{\bf k}[i\omega_n + \mu-\hat{H}^{\mathcal{W}}_{KS}({\bf k})-\hat{\Sigma}+\hat{\Sigma}_{dc}]^{-1}.
\end{aligned}
\end{eqnarray}
In order to enhance convergence, one normally starts with  $\hat{\Sigma}=\hat{\Sigma}_{dc}$, i.e., using the Hartree-energy as a first guess for the self-energy. A total number of \tk-points, $n_{\bf k}$, is considered in the reducible Brillouin Zone. 

(ii) Applying the impurity Dyson equation, which relates the non-interacting impurity Green's function to the (lattice and impurity) self-energy and interacting Green's function
\begin{eqnarray}
\hat{\mathcal{G}}(i\omega_n)^{-1}=\hat{\Sigma}(i\omega_n)+[\hat{G}(i\omega_n)]^{-1}.
\end{eqnarray} 

(iii) Solving the Anderson impurity problem (AIM) defined by the non-interacting Green's function and the local Coulomb interaction $U$, i.e., calculating its interacting Green's function $\hat{{G}}_{imp}(i\omega_n)$
\begin{equation}
\hat{\mathcal{G}}(i\omega_n), U \stackrel{AIM}{\longrightarrow}  \hat{{G}}_{imp}(i\omega_n)
\end{equation}
This is numerically the most involved step; we employ the  continuous-time quantum Monte-Carlo method \cite{CTQMC} in the w2dynamics implementation.\cite{w2d}

(iv) Applying the impurity Dyson equation once again, this time to calculate   the self-energy as the difference between
 the inverse non-interacting impurity Green's function  $\hat{\mathcal{G}}(i\omega_n)$
and the interacting  (lattice and impurity) Green's function $\hat{G}(i\omega_n)$
\begin{eqnarray}
\hat{\Sigma}(i\omega_n) = \hat{\mathcal{G}}^{-1}(i\omega_n)-\hat{G}_{imp}^{-1}(i\omega_n).
\end{eqnarray}
This self-energy is now used again in step (i) to calculate a new local Green's function. This procedure is referred to as ``DMFT cycle'' in Fig.~\ref{cycle}. In a one-shot DFT+DMFT calculation we would stop after this DMFT calculation, extracting the interacting Green's function and further physical quantities from the converged DMFT solution.

By contrast, in CSC DFT+DMFT calculations, we need to determine  the DMFT-modified electron density (lower left segment in  Fig.~\ref{cycle}); recalculate from this the Kohn-Sham potential and the  Bloch waves {\em without DFT self-consistency}; redo for these a Wannier function projection; which is the starting point for  another DMFT step
(see green and blue arrows in Fig.~\ref{cycle}).

We still need to discuss how we calculate the DMFT-modified electron density $\rho({\bf r})= \rho_{DFT}({\bf r})+\Delta \rho({\bf r})$, which we defined in terms of the  Kohn-Sham or DFT $\rho_{DFT}({\bf r})$ and the correlation-induced difference 
$\Delta \rho({\bf r})$. The latter can be calculated as:\cite{Frank}
\begin{equation}\label{diffrho}
\begin{aligned}
\begin{split}
\Delta\rho({\bf r}) &  = \rho_{DMFT}({\bf r})-\rho_{DFT}({\bf r})\\
                            &  = \bra{{\bf r}}(\hat{G}-\hat{G}^{DFT})\ket{{\bf r}}\\
                            &  = \bra{{\bf r}}\hat{G}^{DFT}[\Delta\hat{\Sigma}+(\mu_{DFT}-\mu)]\hat{G}]\ket{{\bf r}}.
\end{split}
\end{aligned}
\end{equation} 
 where $\mu_{DFT}$ and $\mu$ are the DFT and DMFT chemical potentials, respectively and
$\hat{G}^{DFT}(i\omega_n)=\sum_{\bf k}[i\omega_n + \mu_{DFT}-\hat{H}^{\mathcal{W}
}_{KS}({\bf k})]^{-1}$ is the DFT Green's function.
 It is computationally convenient to express $\Delta \rho({\bf r})$ in momentum space, which can be deduced from Eq. (\ref{diffrho}) as
\begin{eqnarray}\label{dn}
\Delta\rho({\bf r})&=&\frac{1}{n_{\bf k}}\sum_{{\bf k},\alpha\alpha'} \braket{{\bf r}|w_{\alpha{\bf k}}}\Delta N^{\mathcal{W}}_{\alpha\alpha'}({\bf k})\braket{w_{\alpha'{\bf k}}|{\bf r}},\\
\Delta \hat{N}^{\mathcal W}({\bf k})&=&\frac{1}{\beta}\sum_n\hat{G}^{DFT}({\bf k},i\omega_n)(\Delta\hat{\Sigma}(i\omega_n)+\Delta\mu)\\\nonumber && \times \hat{G}({\bf k},i\omega_n) 
\end{eqnarray}  
with $\Delta\Sigma=\hat{\Sigma}-\hat{\Sigma}_{dc}$, $\Delta\mu= \mu_{DFT}-\mu$. 
It is to be noted that no convergence factor in the frequency summation needs to be used
 for $\Delta \hat{N}^{\mathcal W}({\bf k})$ because  both Green's functions asymptotically
decay as $1/\omega_n$. Note that the change of occupation in Wannier space  $\Delta N^{\mathcal W}_{\alpha, \alpha'}$ has an explicit ${\bf k}$-dependence, which will have significant consequences in the following section. 

In order to update the DFT charge density we need to
transform  $\Delta \hat{N}^{\mathcal W}({\bf k})$ from the Wannier to the Bloch basis 
using the unitary and disentanglement matrices,   $\hat{U}({\bf k})$ and $\hat{V}({\bf k})$, that define this transformation:
\begin{eqnarray}\label{dn1}
\Delta\hat{N}({\bf k}) &=& \hat{U}({\bf k})\Delta \hat{N}^{\mathcal{W}}({\bf k})\hat{U}^{{\dagger}}({\bf k}) \\ \label{dn2}
\Delta\hat{N}({\bf k}) &=& \hat{V}({{\bf k}})\hat{U}({\bf k})\Delta \hat{N}^{\mathcal{W}}({\bf k}) \hat{U}^{\dagger}({\bf k})\hat{V}^{\dagger}({\bf k})
\end{eqnarray}
Knowing the correlation-induced change of occupation in the Bloch or Kohn-Sham basis we can finally calculate the modified density since we know the
spatial density  $D^{{\bf k}}_{\nu'\nu}({\bf r})=\psi_{{\bf k}\nu}({\bf r})\psi^{*}_{{\bf k}\nu'}({\bf r})$ of each Bloch wave:
\begin{equation}\label{dna}
\Delta \rho({\bf r})=\frac{1}{n_{\bf k}}\sum_{\bf k} \sum_{\nu\nu'=1}^{{\mathcal C}^o}D^{{\bf k}}_{\nu'\nu}({\bf r})\Delta N_{\nu\nu'}({\bf k})
\end{equation}

The full CSC DFT+DMFT   hence consists of the  following workflow,
 schematically  depicted in Fig.~\ref{cycle}:
\begin{itemize}

\item A converged charge density is obtained within DFT to have a reasonable electronic structure to start with (upper left part of Fig.~\ref{cycle}). The target bands are identified as a prelude for the Wannier projection. In the following CSC DMFT cycle (green and blue arrows in Fig.~\ref{cycle}), a single DFT iteration is performed to update the DFT Kohn-Sham Hamiltonian (i.e., without the orange arrow in the upper left part). We employ the Wien2k program package here.

\item Maximally localized Wannier functions  are computed within the target subspace   as explained in Eqs. (\ref{wan0})-(\ref{wan2}) (upper right section of Fig.~\ref{cycle}). The DFT Kohn-Sham Hamiltonian is transformed into the Wannier basis  following Eq.\ (\ref{ham}).  We employ wien2wannier \cite{wien2wannier} and Wannier90 \cite{wanrev} to this end.

\item A single DMFT cycle is performed using  w2dynamics\cite{w2d} (lower right part of Fig.~\ref{cycle}). This provides the self-energy $\hat \Sigma$,  local Green's function $\hat{G}$, and the DMFT chemical potential $\mu$, which is fixed to the particle number. Let us note  that, for practical purposes, it is beneficial to start with a converged ``one-shot" DFT + DMFT calculation. Moreover, a mixing (under-relaxation) between old and new DMFT self-energy is employed.

\item For the correlated charge distribution (lower left part of Fig.~\ref{cycle}), first $\Delta N^{\mathcal W}({\bf k})$ is calculated taking the difference between DMFT and DFT Green's functions,
$\hat{G}$ and  $\hat{G}_{DFT}$, as in Eq.\ (\ref{dn}). As described in Eqs.\ (\ref{dn1})-(\ref{dna}), $\Delta N^{\mathcal W}({\bf k})$ is transformed back to the DFT eigenbasis and used to obtain the correlation-induced change of density $\Delta \rho({\bf r})$ and the
total density  $\rho({\bf r})$ of the correlated solution.

\item The DFT+DMFT charge density, $\rho({\bf r})$, is finally  compared with the old density. If the difference does not satisfy the convergence criteria, the new density is mixed with the old density and the result serves as the new density for a new $V_{DFT}$ and a new solution of the Kohn-Sham equation etc. until convergence.  At the same time, a  convergence of $\hat{G}(\tau)$ is also checked.
\end{itemize}

\section{Applications}
In the following, fully CSC DFT+DMFT calculations are  employed to shed light on correlation-induced charge redistribution beyond the gross effect of moving electrons from a WF centered at one atom to a WF centered at another atom. Two cuprates, Sr$_2$CuTeO$_6$ and HgBa$_2$CuO$_4$, whose physics is dominated by a single band, are studied. The systems are different in several aspects. First Sr$_2$CuTeO$_6$ exhibits a single isolated band around Fermi-energy, while in HgBa$_2$CuO$_4$ the single $d$ band is entangled with other bands crossing it. 
On the technical side this requires disentanglement to project onto a single Wannier $d$ orbital for  HgBa$_2$CuO$_4$ as discussed in the previous section.

Next, a multi-orbital situation is considered with a single, free-standing layer of SrVO$_3$ and  $t_{2g}$ orbitals at the Fermi energy that are well isolated from the other orbitals. Here, the interplay between structural confinement, orbital ordering, electronic correlations and CSC  is discussed in detail. 

\subsection{Sr$_2$CuTeO$_6$}
\label{sec:SCTO}
\begin{figure}[tb]
\includegraphics[width=0.5\textwidth]{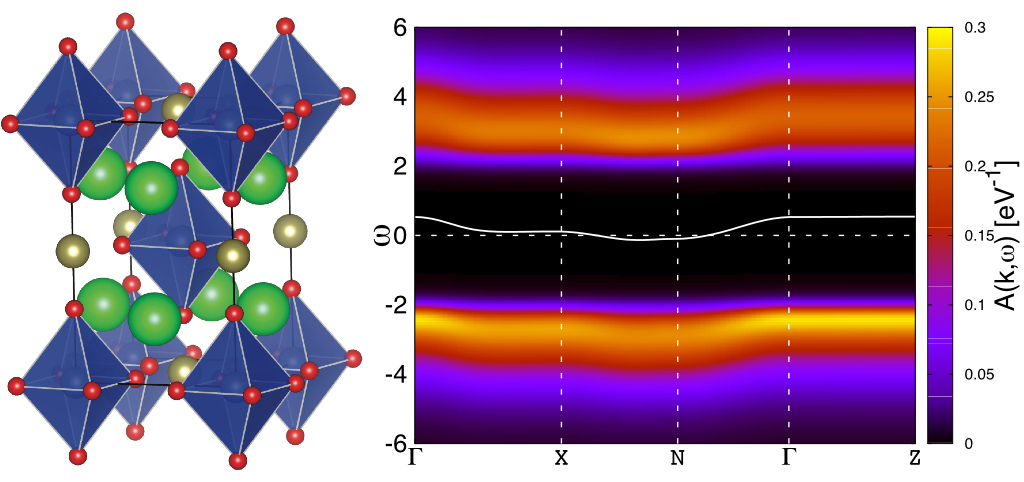}
\centering
(a) \hskip 3.0cm (b) \\~\\
\caption{\label{fig1} (Color online) (a) Crystal structure of Sr$_2$CuTeO$_6$ with green, golden, and red balls representing  Sr, Te, and O, respectively; the Cu sites are in the center of the blue octahedra that are elongates along the $c$-axis. (b) Spectral function, $A({\bf k},w)$, as calculated by DFT+DMFT along a high symmetry path through the Brillouin zone. The white curve depicts well separated wannier band; the Fermi energy is set at zero.}
\end{figure}

 To describe the physics of cuprates, an effective single band model can be derived where the contributing orbital is predominantly of Cu-$d_{x^2-y^2}$ character, with some admixture of O-$p_{x/y}$. The compound, Sr$_2$CuTeO$_6$, exhibits square lattice Heisenberg antiferromagnetism\cite{scto_expt} in a quasi two-dimensional plane, consisting of  Cu and O atoms, see Fig.\ \ref{fig1} (left).  It is quite unique in the cuprate group in having a completely isolated and weakly dispersing  band around the Fermi-energy, see the white band in  Fig.\ \ref{fig1} (right).  Thus, no disentanglement is needed in this material.

We take for our calculations the $I4/m$ symmetry of the lattice with the experimental lattice  parameters\cite{struct}, i.e., in-plane lattice constant a= 5.4308 \AA, out of plane lattice constant c = 8.4664 \AA. A slight complication of the lattice structure is that the CuO$_6$ octahedra in  Sr$_2$CuTeO$_6$  are rotated around the $z$-direction. In contrast to the CuO$_2$ planes of other cuprates, cf.\ Section \ref{sec:HBCO} below, Sr$_2$CuTeO$_6$ has planes with four O per Cu; no oxygen is shared, which explains the low  itinerancy.  

In DFT, Sr$_2$CuTeO$_6$ is a metal with a single half-filled band crossing the Fermi energy, predominantly of Cu-$d_{x^2-y^2}$ character.  Electronic correlations result, however, in an insulating phase with two Hubbard bands separated by $U$.  This is captured in our DMFT calculations, performed with $U = 6.5\,\text{eV}$  at inverse temperature $\beta = 40\,\text{eV}$. During the charge self-consistency cycle, the self-energy and density are under-relaxed; 1000 k-points  are considered in the reducible Brillouin zone for all the calculations.

The half-filled DFT band remains half-filled in DMFT.
Naively one might expect that for an unchanged $d$-electron occupation (half filling) there can be no CSC effect.
However, the occupation in  \tk-space is altered. In DFT (white band in  Fig.\ \ref{fig1}) some \tk-points (in-between $X$-$N$-$\Gamma$) are below the Fermi level, and hence filled with one electron, whereas for all other \tk-points the occupation is zero.

 In DMFT this half-filled band is split into two Hubbard bands that are broadened 
because of the imaginary part of the self-energy, the lifetime. This splitting means that now every $\mathbf  k$-point 
 is occupied with half an electron (lower Hubbard band), whereas the remaining half electronic state (upper Hubbard band) remains unoccupied.
That is, we have a major change of the occupation $\Delta N({\mathbf k})$
in  $\mathbf  k$-space, as  calculated from the differences between $G$ and $G_{DFT}$ at each ${\bf k}$-point in Eq. (\ref{dn}). For the orbital occupation the sum 
over the entire Brillouin zone is taken, preserving the number of electrons in the $d_{x^2-y^2}$ orbital.

For the change of charge $\Delta \rho({\mathbf r})$ in real space, however, each  $\Delta N({\mathbf k})$ in Eq. (\ref{dna}) is weighted with the spatial distribution of the corresponding Wannier functions. Hence, the splitting into  Hubbard bands results 
in a charge redistribution: the Wannier functions have a different spatial dependence at each \tk-point. 
\begin{figure}[H]
\centering
\includegraphics[width=0.3\textwidth] {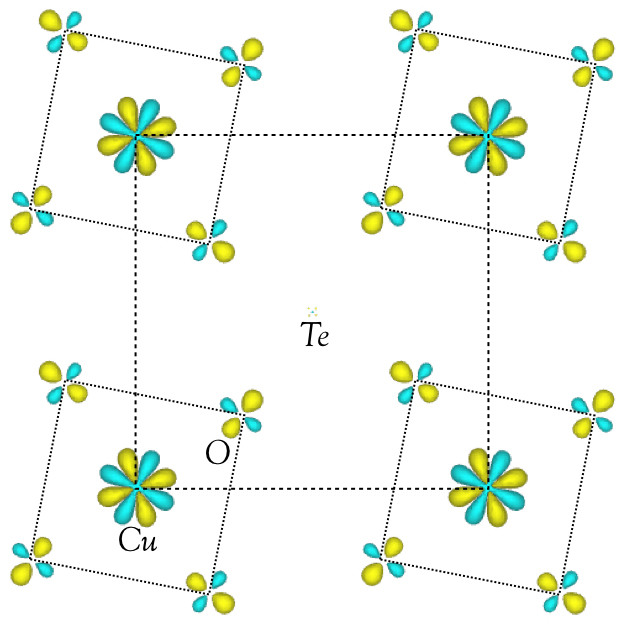}
\caption{\label{rdel} Iso-surface plot of the correlation-induced charge density difference, $\Delta \rho({\bf r})=\rho_{\rm DMFT}({\bf r})-\rho_{\rm DFT}({\bf r})$, in Sr$_2$CuTeO$_6$. Yellow and cyan correspond to positive and negative $\Delta \rho({\bf r})$'s at an iso value of 4$\times$10$^{-3}$ electrons/bohr$^3$; the dashed  and dotted lines represent the unit cell and the (rotated) four O atoms around each Cu site, respectively.}
\end{figure}
This correlation-induced correction to the charge density within the Cu-O plane is shown in Fig.  \ref{rdel}. Here, the yellow (cyan) color corresponds to a gain (loss) of electron density in real space. As the single band of our consideration is predominantly of $d_{x^2-y^2}$ character, the contribution of each sign has the same orbital symmetry; the total change in density within the unit cell (shown as the black dashed box) is zero.  The charge redistribution around each Cu ion can be understood easily for cubic crystal symmetry; see for HgBa$_2$CuO$_4$ in section \ref{sec:HBCO}.  Here, with lower symmetry, the charge redistribution  around each Cu ion shows eight lobes with positive and negative contributions. Each Cu is surrounded by 4 oxygen atoms at the edges of the 
dotted box. As one can clearly see the positive contribution at these O sites is larger than the negative one. That means that even in our $d$-only model calculation there is some charge redistribution from Cu $d$ to oxygen $p$. This is akin to the situation in $d$-$p$ models where charge is moved from $d$ to $p$ orbitals as well. However in our calculation this effect 
occurrs even though we have only  a single orbital in the DMFT calculation. This Wannier orbital is centered around the Cu sites and predominantly  of $d_{x^2-y^2}$ character. But it has some admixture of oxygen $p$, i.e., it has some charge density  at the neighbouring oxygen sites as well. 
This admixture requires some \tk-dependence of the Wannier functions [Eq.~(\ref{wan2})]; and the occupation is reduced, \eg, around  the $N$ point, while increased in the remainder, eventually leading to the charge distribution pattern of Fig. \ref{rdel}.
\subsection{HgBa$_2$CuO$_4$}
\label{sec:HBCO}
\begin{figure}[H]
\centering
\includegraphics[width=0.5\textwidth]{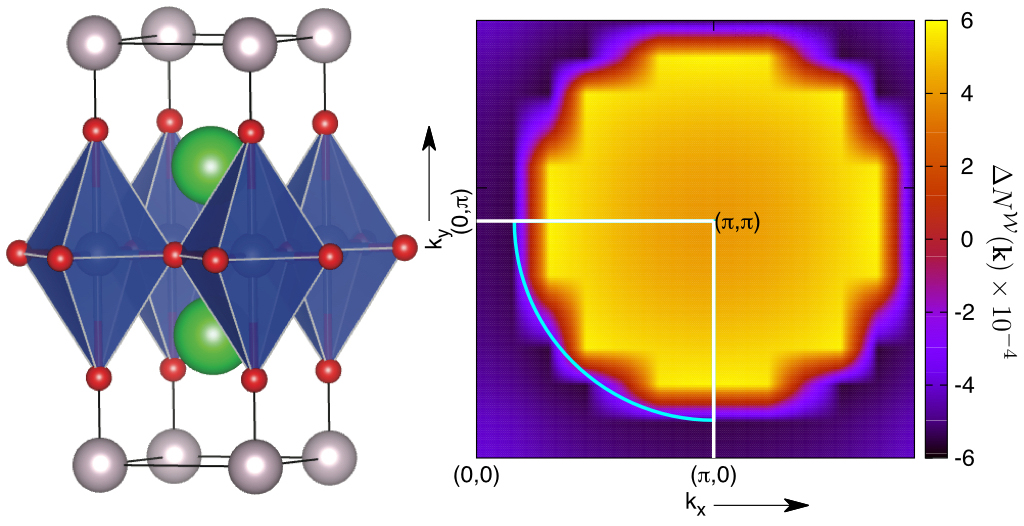}
(a) \hskip 3.5cm (b) \\~\\
\caption{\label{hbcodn} (Color online) (a) Crystal structure of HgBa$_2$CuO$_4$. Green, pink, blue and red balls represent Ba, Hg, Cu and  O atoms, respectively. (b) {\bf k}-dependence of $\Delta N^{\mathcal W}$, i.e., the change of occupation of the Wannier orbitals in \tk-space for the $k_z = 0$ plane in the Brillouin zone. The high-symmetry points $\Gamma\ (0,0)$, $X\ (\pi,0)$, and $M\ (\pi,\pi)$ are marked.  The cyan curve depicts the DFT Fermi surface which separates positive (yellow) and negative (blue) $\Delta N^{\mathcal W}$ contributions.  }
\end{figure}

Let us now turn to HgBa$_2$CuO$_4$, a prototype of the high temperature cuprate superconductors.\cite{HBCO}
The arrangement of the CuO$_6$ octahedra is distinctly different in undoped HgBa$_2$CuO$_4$  compared to that of Sr$_2$CuTeO$_6$, see the crystal structure in
 Fig.~\ref{hbcodn}(a). The system belongs to the space group P4/mmm, with Hg, BaO, CuO$_2$, and BaO layers stacked vertically along the $c$-axis of a tetragonal unit cell. Each oxygen atom in the CuO$_2$ plane is shared by two Cu atoms, resulting in a more direct hopping and a larger bandwidth of the Cu-$d_{x^2-y^2}$ band compared to that of  Sr$_2$CuTeO$_6$. 
But 
the  $d_{x^2-y^2}$  bands of  HgBa$_2$CuO$_4$ are no longer isolated. This requires disentanglement for  constructing  an effective single band model.

\begin{figure}[H]
\centering
\includegraphics[width= 0.49\textwidth] {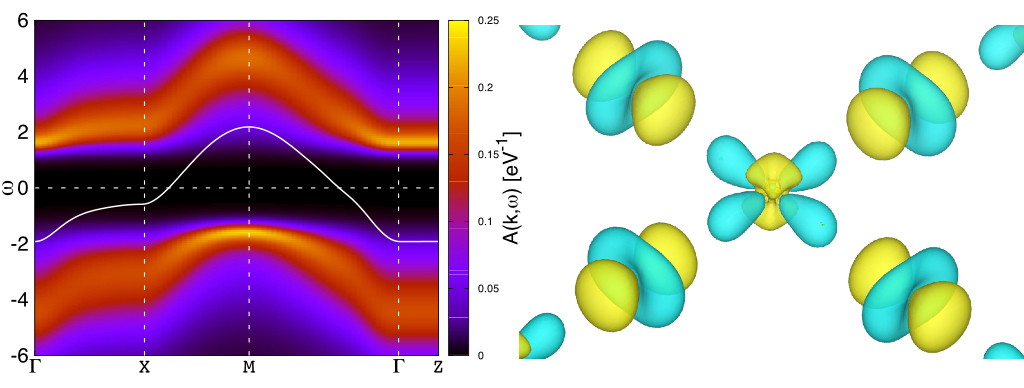}
(a) \hskip 3.5cm (b) \\~\\
\caption{\label{hbcoakw}  (Color online) (a) {\bf k}-resolved DMFT
 spectral function $A({\mathbf k}, \omega)$  (color) in comparison with the
 DFT bandstructure (white line) for HgBa$_2$CuO$_4$.  The dashed horizontal line is the Fermi energy.  (b) Iso-surface plot of the DMFT charge redistribution $\Delta \rho({\bf r})$.  Yellow and cyan correspond to positive and negative values of $\Delta \rho({\bf r})$, at an iso value of 1.5$\times$10$^{-3}$ electrons/bohr$^3$).
}
\end{figure}

Like Sr$_2$CuTeO$_6$, HgBa$_2$CuO$_4$ is metallic in DFT, but is insulating if electron correlations are included as well as in experiment. For all the calculations
we use  845 ${\bf k}$-points in the full Brillouin zone; and 
for the  DMFT  at $\beta$ = 40 we employ $U$= 6.5 eV.
This splits the DFT band into two Hubbard bands  (see Fig.\ \ref{hbcoakw}) 
and redistributes
the  ${\bf k}$-space occupation of the Wannier orbitals as in the case  of Sr$_2$CuTeO$_6$ [Fig.\ \ref{fig1} (a)].
The most remarkable difference to that material is the  much larger bandwidth in both DFT (white line) and DMFT (color).

Hence, for the very same reason as in the previous Section,
$\Delta N^{\mathcal W}$  has a strong ${\bf k}$-dependence.
Since we have an  effectively two-dimensional model, we plot
 in Fig.~\ref{hbcodn}(b) $\Delta N^{\mathcal W}({\bf k})$ in the plane $k_z=0$ of the Brillouin zone.
  The yellow section of the plane represents the set of ${\bf k}$-points that have positive $\Delta N^{\mathcal W}$. These states were  unoccupied in  DFT but get half-occupied in DMFT due to the lower Hubbard band dispersing throughout the Brillouin zone. The  negative counterpart (blue) is around the $\Gamma$ point where all states where occupied in DFT.  The boundary between  these two regions is exactly the DFT Fermi-surface marked with a cyan line.

In spite of the fact that both cuprates can be modeled using only a single band, there is a significant correction to the  charge density  by DMFT.  In Fig.~\ref{hbcoakw}(b), the correlation induced  charge redistribution $\Delta \rho({\bf r})$ is depicted as an iso-surface plot. A negative sign of $\Delta \rho({\bf r})$ (cyan) suggests electron loss from the single band comprised of Cu-$d_{x^2-y^2}$ and O-$p$ orbitals, a positive sign (yellow) electron gain. 
There are gains as well as losses around both  Cu and O sites; altogether   some net charge is transferred from Cu to O. 

Let us hence focus on the O charge gain and Cu loss in the following.
Around the O sites the gain has the form of a $p_x$- and $p_y$-orbital density 
 pointing to the Cu sites. It stems  from the admixture of these orbitals to our single band. On the Cu site in turn there are blue  $d_{x^2-y^2}$-like lobes of removed charge pointing towards the neighboring oxygen sites. This indicates that the redistribution  $\Delta N^{\mathcal W}$ of Wannier orbitals in \tk-space, effectively reduces the level  of admixture between these orbitals  when moving from the DFT metal to the DMFT insulator.  Even though one might naively assume that in a single Wannier band with fixed occupation CSC effects are minor,  the density correction is significant for both cuprates. 

\subsection{SrVO$_3$}
\label{sec:SVO}
SrVO$_3$ crystallizes in a cubic perovskite lattice structure and has been the testbed material for DFT+DMFT \cite{SrVO3exp,Pavarini03,Liebsch03a,Nekrasov05b,Nekrasov05a} and GW+DMFT \cite{P7:Casula12b,Tomczak12,Taranto13,Tomczak14} method development. A strong interplay between the octahedral crystal field in VO$_6$ and electron correlation determines the properties of this material. Bulk SrVO$_3$ exhibits robust metallicity also upon chemical substitution such as  in Ca$_{1-x}$Sr$_x$VO$_3$\cite{bulksvo}. However, the material undergoes a metal-insulator transition if  manipulating its dimensionality\cite{svo_expt}. Ultra thin layers (up to 3 monolayers) of SrVO$_3$ are insulating, which opens the  possibility to control the metal-insulator transition by applied electric field or strain, paving the way for a Mott transistor\cite{zhong}.

In ultra-thin layers the bulk $t_{2g}$ symmetry is broken:  the out-of-plane $d_{xz/yz}$ orbitals have a reduced bandwidth, while the in-plane $d_{xy}$ bandwidth remains almost unchanged. Given the 3$d^1$ electronic configuration of vanadium, ultra-thin layers hence favor the electrons to be placed in the $d_{xy}$ orbital. An orbital polarization develops. 
The orbital reoccupation is quite dramatic in DFT+DMFT: from  1/3 for all  $t_{2g}$ Wannier orbitals in the metallic bulk to almost  a occupation of 1 electron  in the $d_{xy}$ orbital for a on-layer film. Let us note that DFT underestimates the orbital polarization which is strongly enhanced by electronic correlations: in DFT $d_{xy}$ and $d_{xz/yz}$ orbitals have 0.6 and 0.2 electrons, respectively, for the ultra-thin film; and it is metallic.
Hence  a freestanding monolayer of SrVO$_3$ is ideally suited to study the CSC DFT+DMFT charge redistribution caused by  an orbital polarization.
\begin{figure}[H]
\centering
\includegraphics[width= 8.0cm] {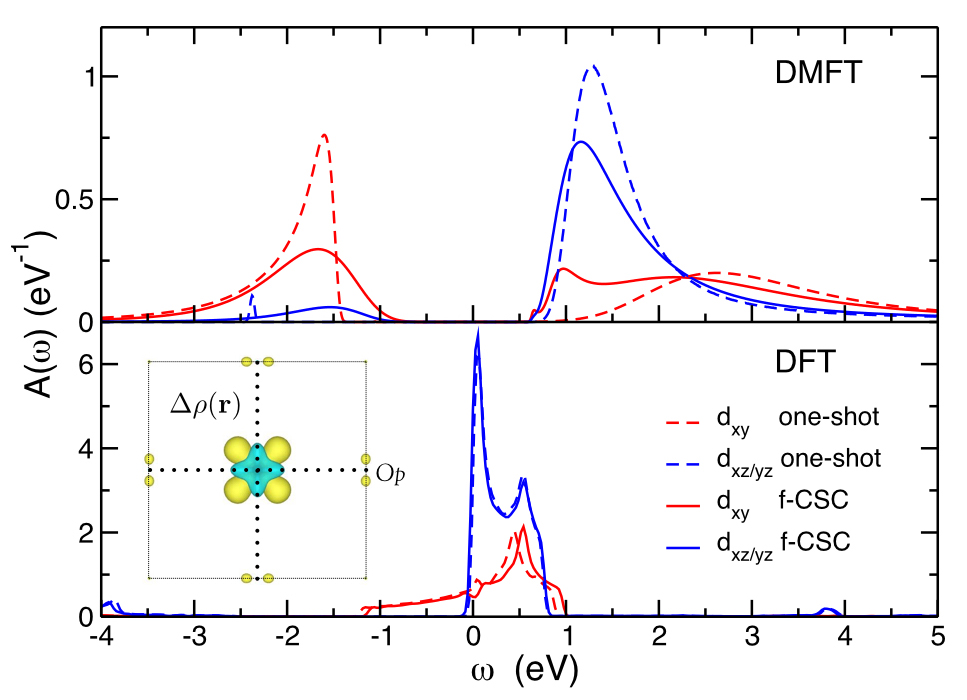}
\caption{\label{dos} (Color online) DMFT (upper panel) and DFT (lower panel) density of states projected onto the V-$t_{2g}$ orbitals. Dashed lines correspond to calculations without CSC; solid lines are with CSC. Inset: Top view of the iso-surface plot of charge redistribution, $\Delta \rho({\bf r})$.  Yellow and cyan  correspond to positive and negative contributions at an iso value of 6$\times$10$^{-3}$ electrons/bohr$^3$.}
\end{figure}

In our calculation we set the lattice constant (a=3.92\AA) 
to that of a single  SrVO$_3$ layer on SrTiO$_3$ as this is the experimental substrate\cite{svo_expt}.
 Fig.\ \ref{dos} (lower panel)  shows the DFT density of states (DOS); and the dashed lines represent the DOS of the V-$t_{2g}$ orbitals in DFT.
 Let us note that the position and width of the $d_{xy}$ band (red) is significantly larger compared to that of the $d_{xz/yz}$ bands (blue). This leads to a first tendency towards an orbital polarization already in DFT, giving the aforementioned occupations.

For the DMFT calculations at inverse temperature $\beta=40$, we employ the Kanamori interaction parameters $U'=4.0\,$eV, $J=0.75\,$eV from the literature.\cite{zhong} The effect of electron correlations in Fig.~\ref{dos} (upper panel)  is twofold: (i) the system becomes a Mott insulator, and (ii) in the insulating phase, $d_{xy}$ is half-filled while the other orbitals are essentially empty.  This kind of physics has been observed before\cite{zhong} but let us now turn to the effect of CSC.

The dashed curves represent the  spectral function corresponding to the one-shot DFT+DMFT calculation without CSC, while the solid curves represent the full CSC results. Let us note that the insulating energy gap is slightly reduced by the CSC, which can be attributed to the charge redistribution within the $t_{2g}$ manifold shown as an inset in  Fig.\ \ref{dos}. As is to be expected, the charge redistribution has a positive $d_{xy}$-like shape as these orbitals become more occupied. Perpendicular to the plane, shown in the inset, there is a reduced charge in a $d_{xz}$- and  $d_{yz}$-like shape. 

This changed orbital occupation influences, in turn, the DFT electronic structure, see solid curves in  Fig.\ \ref{dos} (lower panel): The $d_{xy}$ orbital  that is more occupied in DMFT, is shifted upwards to higher energies in DFT and vice versa for $d_{xz/yz}$, as is to be expected already from the Hartree term. 
That is, DFT partially compensates the correlation effect of DMFT, but a large net effect remains. 
This net effect is shown in  Fig.~\ref{dos}: the charge redistribution in the inset, the CSC DFT and DMFT results as solid lines in the main panel. One shot DFT+DMFT has almost filled $d_{xy}$ orbital and almost empty $d_{xz}$- and  $d_{yz}$-orbitals, while full CSC results in a slight reduction of $d_{xy}$-orbital occupancy, compared to that in one-shot calculation and vice versa for $d_{xz}$- and  $d_{yz}$-orbitals. The effect of full CSC is not negligible, also regarding the reduction of the band-gap.

\section{Summary and conclusions}
\label{sec:summary}
We have implemented a fully charge self-consistent DFT+DMFT method, using maximally localized Wannier functions constructed with Wannier90 \cite{wanrev}, the Wien2k program package, wien2wannier as an interface, and w2dynamcis as an impurity solver.  We applied the method to strongly correlated electron systems and discussed different physical and technical aspects.

The  cuprates, Sr$_2$CuTeO$_6$ and HgBa$_2$CuO$_4$, can be modeled by a single Wannier orbital. In this situation, one might assume that the charge self-consistency has no effect since this single orbital must remain half-filled; there is no charge redistribution to other orbitals. Nonetheless, the real space charge density is  changed with full CSC DFT+DMFT. 

In both cuprates, charge is removed from around the Cu site and added around the O sites. Note that oxygen $p$-states are mixed into the single, predominantly  $d_{x^2-y^2}$ orbital.
The reason for this change is a change of occupation of the Wannier orbitals in \tk-space. While for  the metallic DFT solution,  Wannier functions in some part of the the Brillouin Zone are singly occupied, in DMFT the band splits into two Hubbard bands and all  \tk-points are occupied equally with half an electron.

Besides this common ground, there are some differences between Sr$_2$CuTeO$_6$ and HgBa$_2$CuO$_4$. The former has much weaker $p$-$d$ hybridization and itinerancy. Hence we do not need disentanglement to Wannier project onto the single orbital. As for the CSC, the changes at the oxygen are much less pronounced because the O states admix to a much lesser extent in  Sr$_2$CuTeO$_6$. It also has a lower symmetry, which results in a more complicated charge redistribution pattern.

 A significant correlation-induced occupation redistribution within  the V-$t_{2g}$ manifold is observed in a single layer of SrVO$_3$. Here, the interplay between crystal field and electron correlation results in a pronounced orbital polarization. The orbital polarization can be clearly identified in the charge redistribution. The CSC has the tendency to counteract the DMFT orbital polarization, which is however hardly reduced at self-consistency with respect to one-shot DFT+DMFT.

 In all the cases, using single or multi-band models, $\Delta N^{\mathcal W}({\bf k})$ has a significant ${\bf k}$-dependence which translates to an ${\bf r}$-dependence of $\Delta \rho({\bf r})$.  This shows that there are more profound effects of CSC in DFT+DMFT than the gross effect of charge redistribution from one site to another found in previous studies.

\section*{Acknowledgement}
We thank Patrik Thunstr\"{o}m, Rainer Bachleitner, Markus Wallerberger, Oleg Janson and Peter Blaha for valuable discussions.  Financial support from the European Research Council under the European Union's Seventh Framework Program (FP/2007-2013)/ERC through grant agreement n.\ 306447 and by the Austrian Science Fund (FWF) through SFB ViCoM project ID F4103 and I 1395, which is part of the DFG research unit FOR 1346, as well as START project Y746 is gratefully acknowledged.  Calculations were done in part on the Vienna Scientific Cluster (VSC).


\end{document}